\documentclass[superscriptaddress,aps,twocolumn,preprintnumbers,showpacs,showkeys,nofootinbib,floatfix]{revtex4-1}

\usepackage{amssymb,amsmath,amsfonts,amsthm,graphicx,psfrag}
\newcommand{\beq}{\begin{eqnarray}}
\newcommand{\eeq}{\end{eqnarray}}

\begin{document}
\preprint{}

\title{Monte-Carlo study of Dirac semimetals phase diagram.}

\author{V.~V.~Braguta}
\email[]{braguta@itep.ru}
\affiliation{Institute for High Energy Physics NRC "Kurchatov Institute", Protvino, 142281 Russian Federation}
\affiliation{Institute of Theoretical and Experimental Physics, 117259 Moscow, Russia}
\affiliation{Far Eastern Federal University, School of Biomedicine, 690950 Vladivostok, Russia}
\affiliation{Moscow Institute of Physics and Technology, Institutskii per. 9, Dolgoprudny, Moscow Region, 141700 Russia}

\author{M.~I.~Katsnelson}
\email[]{m.katsnelson@science.ru.nl}
\affiliation{Radboud University, Institute for Molecules and Materials,
Heyendaalseweg 135, NL-6525AJ Nijmegen, The Netherlands}
\affiliation{Ural Federal University, Theoretical Physics and Applied Mathematics Department, Mira Str. 19, 620002 Ekaterinburg, Russia}

\author{A.~Yu.~Kotov}
\email[]{kotov@itep.ru}
\affiliation{Institute of Theoretical and Experimental Physics, 117259 Moscow, Russia}
\affiliation{National Research Nuclear University MEPhI (Moscow Engineering Physics Institute), Kashirskoe Highway, 31, Moscow 115409, Russia}

\author{A.~A.~Nikolaev}
\email[]{nikolaev.aa@dvfu.ru}
\affiliation{Institute of Theoretical and Experimental Physics, 117259 Moscow, Russia}
\affiliation{Far Eastern Federal University, School of Biomedicine, 690950 Vladivostok, Russia}

\begin{abstract}
In this paper the phase diagram of Dirac semimetals is studied within lattice Monte-Carlo simulation.
In particular, we concentrate on the dynamical chiral symmetry breaking which results in semimetal/insulator
transition. Using numerical simulation  we determined the values of the
critical coupling constant of the semimetal/insulator transition for different values of the anisotropy of the Fermi velocity.
This measurement allowed us to draw tentative phase diagram for Dirac semimetals. 
It turns out that within the Dirac model with Coulomb interaction both
Na$_3$Bi and Cd$_3$As$_2$ known experimentally to be Dirac semimetals
would lie deeply in the insulating region of the phase diagram. It probably
shows a decisive role of screening of the interelectron interaction in real
materials, similar to the situation in graphene.
\end{abstract}



\maketitle

{\bf Introduction.}
Recent significant advances in condensed matter physics are connected to the discovery
of new materials with remarkable properties. Probably the discovery of graphene\cite{Novoselov666, Geim2007}
is the most famous example.  Graphene attracts considerable interest because of its unique electronic properties; most of them are related to existence of two conical points in the electron energy spectrum (Fermi points) and massless fermion excitations
which are similar to 2D Dirac fermions \cite{PhysRev.71.622,PhysRev.104.666,Semenoff:1984dq,Novoselov2005,Zhang2005}.

Lately there was theoretically predicted\cite{PhysRevB.85.195320,PhysRevB.88.125427} and shortly afterwards experimentally confirmed the existence of so-called Dirac semimetals
Na$_3$Bi\cite{Liu864} and Cd$_3$As$_2$\cite{Neupane2014,PhysRevLett.113.027603} which manifest the properties of 3D analog of graphene.
Low energy spectrum of these materials is determined by two Fermi points. In the vicinity of each Fermi point
the fermion excitations reveal the properties of massless 3D Dirac fermions with the dispersal relation
\begin{equation}
E^2=v^2_{\parallel}(k^2_x+k_y^2)+v^2_{\perp}k^2_z,
\label{eq:dispersion}
\end{equation}
where $v_{\parallel}, v_{\perp}$ are Fermi velocities in the $(x,y)$ plane and $z$ direction correspondingly.
For the Na$_3$Bi: $v_{\parallel}/c\simeq0.001, v_{\perp}/v_{\parallel}\simeq0.1$\cite{Liu864} and for the Cd$_3$As$_2$:
$v_{\parallel}/c \simeq0.004, v_{\perp}/v_{\parallel}\simeq0.25$\cite{Liu2014}.

Due to the smallness of the Fermi velocities magnetic interactions and retardation
effects can be safely disregarded. As the result the interaction in Dirac semimetals is reduced to instantaneous
Coulomb potential with the effective coupling constant $\alpha_{eff}=\alpha_{el} \cdot c / v_{\parallel} > 1$,
where $\alpha_{el}=1/137$. So one sees that the interaction is quite strong
what can dramatically modify properties of these materials. In particular, it
is known that strong interaction between quasiparticles can lead to dynamical chiral symmetry
breaking, formation of energy gap in the fermion spectrum and transition from
semimetal to insulator phase.

This paper is devoted to the investigation of the phase diagram of Dirac semimetals.
In particular, we are going to study semimetal/insulator phase transition in the parameters plane
$(\alpha_{eff}, v_{\perp}/v_{\parallel})$ which results from dynamical chiral symmetry breaking at sufficiently strong
interactions between quasiparticles. To carry out this study we are
going to use lattice Monte-Carlo simulation which fully accounts many-body effects in Dirac semimetals
for arbitrary coupling constant $\alpha_{eff}$. This approach proved to be
very efficient in studying the properties of the strongly correlated systems, for instance,
graphene \cite{Drut:2008rg, Ulybyshev:2013swa, Boyda:2016emg}.
It should be noted that earlier the phase diagram of Dirac semimetals was studied within mean field \cite{Sekine:2014yna, Araki:2015php},
renormalization group \cite{Gonzalez:2015tsa}, Dyson-Schwinger equation \cite{Gonzalez:2015iba}.

Taken into account the spectrum of low energy fermion excitations near the Fermi points and the properties of the interactions discussed above,
the partition function of Dirac semimetals can be written in the following form
\beq
Z=\int D\psi~ D\bar\psi~DA_4~\exp{\bigl ( -S_E \bigr )},
\label{Z}
\eeq
where $\bar \psi, \psi$ are fermion fields, $A_4$ is temporal component of the vector potential of the electromagnetic field.
The Euclidean action $S_E$ can be written as
\begin{equation}
\begin{split}
    S_E=\sum\limits_{a=1}^{N_f=2}\int d^3x dt \bar{\psi}_a (\gamma_4(\partial_4+iA_4)+\xi_i\gamma_i\partial_i)\psi_a+\\
    +\frac{1}{8\pi\alpha_{eff}}\int d^3x dt (\partial_i A_4)^2
\label{eq:continuousaction}
\end{split}
\end{equation}

Here $\gamma_1,\ldots,\gamma_4$ are Euclidean gamma matrices: $\{\gamma_{\mu},\gamma_{\nu}\}=2\delta_{\mu,\nu}$,  and $\xi_i$ are factors, which take into account the anisotropy of the Fermi velocity ($\xi_1=\xi_2=1, \xi_3=v_{\perp}/v_{\parallel}$).

In Eq.~(\ref{eq:continuousaction}) we rescaled $t$ and $A_4$, what allowed to reabsorb the Fermi velocity $v_{\parallel}$
by $\alpha_{eff}$.
As was noted above the smallness of the Fermi velocity $v_{\parallel}\ll  c$ leads to the fact that the
interaction between quasi-particles is instantaneous Coulomb, which is transmitted by the field $A_4$.
Partition function (\ref{Z}) doesn't depend on the vector part of the gauge potential $A_i$ since we are working
at the leading approximation in $v_{\parallel}$.

{\bf Lattice field theory for Dirac semimetals.}
In lattice Monte-Carlo approach one discretizes the continuum expression for the action (\ref{eq:continuousaction}).
In our simulations we use staggered discretization for fermions\cite{Montvay:1994cy}, coupled to Abelian lattice gauge field $\theta_4(x)$:
\begin{equation}
\begin{split}
S_f=\bar{\Psi}_xD_{x,y}\Psi_y=\sum\limits_x\left(ma\bar{\Psi}_x\Psi_x+\right.\\
\left.
+\frac12[\bar{\Psi}_x\eta_4(x)e^{i\theta_4(x)}\Psi_{x+\hat{4}}-\bar{\Psi}_{x+4}\eta_4(x)e^{-i\theta_4(x)}\Psi_x]\right.+\\
+\frac12\sum\limits_{i=1}^3\xi_i[\bar{\Psi}_x\eta_i(x)\Psi_{x+\hat{\i}}-\bar{\Psi}_{x+\i}\eta_i(x)\Psi_x]
\left.\right),
\label{eq:staggered}
\end{split}
\end{equation}
where $\eta_{\mu}(x)=(-1)^{x_0+\ldots+x_{\mu-1}},\mu=1,\ldots,4$ are staggered factors corresponding to $\gamma$-matrices.
The lattice field $\theta_4$ is related to the continuum Abelian field $A_4$ as $\theta_4=a A_4$, where $a$ is a lattice spacing.
It should be noted that nonzero mass term in (\ref{eq:staggered}) is
necessary in order to ensure the invertibility of the staggered Dirac operator $D_{x,y}$. Physical results for zero mass
are obtained by extrapolation of the expectation values of physical observables to the limit 
$m\to0$\footnote{In this paper we express all dimensional observables in lattice units.}.

For discretization of the Abelian field the noncompact action was used:
\begin{equation}
S_g=\frac{\beta}2\sum_{x,i}(\theta_4(x)-\theta_4(x+i))^2.
\label{eq:noncompactqed}
\end{equation}

Here the constant $\beta$ is given by the formula $\beta=\frac{1}{4\pi\alpha_{eff}}$.

Integrating out fermion degrees of freedom one gets the following expression for the
partition function
\beq
\nonumber
Z=\int D\theta_4(x) \exp{\bigl ( -S_{eff} \bigr )}, \\
S^{(eff)}=-\ln\det D[\theta]+S_{g}.
\label{eq:effaction_4}
\eeq
Notice, however, that effective action (\ref{eq:effaction_4}) in continuum corresponds
to four degenerate fermion flavours\cite{Montvay:1994cy} instead of two ones
observed in Na$_3$Bi and Cd$_3$As$_2$. In order to get two fermion flavours
we take square root from the determinant of the Dirac operator what in
the numerical simulation is realized through the rooting procedure.
Thus the effective action used in the simulation is
\beq
S^{(eff)}=-\frac{1}{2}\ln\det D[\theta]+S_{g}.
\label{eq:effaction}
\eeq
For generation of the field $\theta_4(x)$ with the statistical weight $\exp(-S^{(eff)}[\theta])$ the standard Hybrid Monte-Carlo Method\cite{Montvay:1994cy} was used.

\begin{figure}[t]
\centering
\includegraphics[width=9.0cm,angle=0]{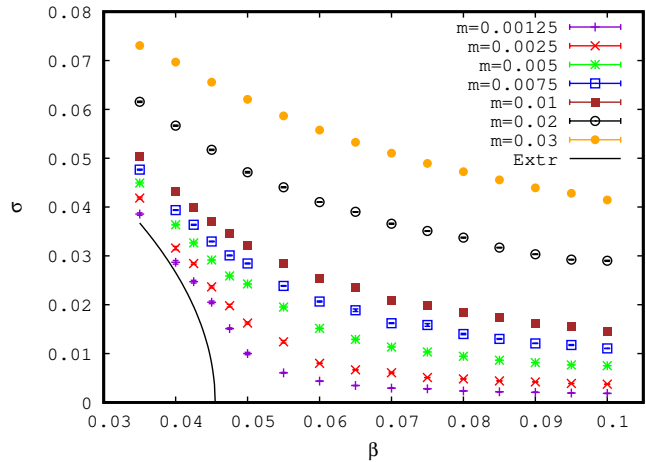}
\caption{The chiral condensate $\langle\bar{\Psi}\Psi\rangle$ as a function of $\beta$ for different values of mass $m$. Black line corresponds to chiral limit $m\to0$ taken with the help of EoS~(\ref{eq:eos}).}
\label{fig:sigmaiso}
\end{figure}

As was noted above  we are going to study semimetal/insulator phase transition
which is connected to dynamical chiral symmetry breaking. To determine the position
of the phase transition we are going to measure the order parameter of chiral symmetry breaking
-- the chiral condensate $\sigma=\langle\bar{\Psi}\Psi\rangle$. In the chiral limit $m=0$;
 $\sigma=0$ in the chiral symmetric phase and $\sigma\neq0$ in the phase where
chiral symmetry is broken.

In addition to the chiral condensate we will calculate
the susceptibility of the chiral condensate $\chi_L=\frac{\partial\sigma}{\partial m}$.
The observable related to the susceptibility and sensitive to the semimetal/insulator phase transition is the logarithmic derivative of the chiral condensate $R=\frac{\partial \ln \sigma}{\partial \ln m}$. In the chiral limit the $R$ reveals the following properties:
in the chirally symmetric phase $\sigma \sim m$ and $R \to 1$.
At the critical point $R \to 1/\delta$, where $\delta$ is a
universal critical exponent and $R \to 0$ in the phase with
broken chiral symmetry.

{\bf Numerical results.} First let us study the case without anisotropy of Fermi velocity in different directions ($\xi_1=\xi_2=\xi_3=1$).
In numerical simulation of Dirac semimetals we used lattice $20^4$.
In Fig.~\ref{fig:sigmaiso} the dependence of $\sigma$ on $\beta$ for different fermion masses is presented.
It is seen from this plot that the formation of the chiral condensate takes place at values of $\beta<\beta_c$
with critical value $\beta_c\sim0.04-0.06$.

\begin{figure}[t]
\centering
\includegraphics[width=9.0cm,angle=0]{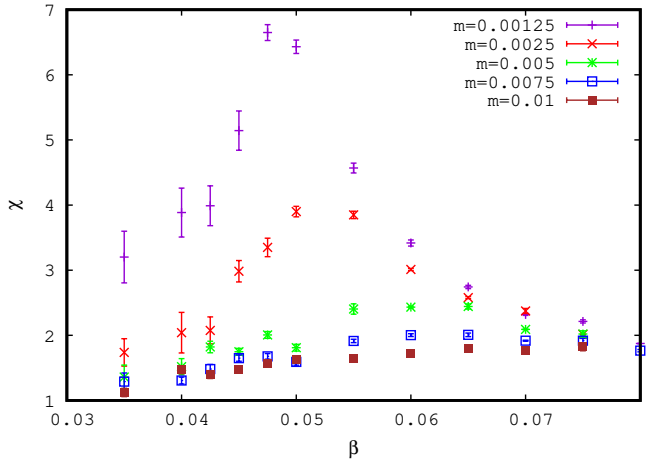}
\caption{Susceptibility $\chi$ of the chiral condensate as a function of $\beta$ for different values of mass $m$.}
\label{fig:susiso}
\end{figure}

In order confirm this result we also studied the susceptibility of the chiral condensate $\chi_L=\frac{\partial\sigma}{\partial m}$ (Fig.~\ref{fig:susiso}) as a function of $\beta$ for different values of mass. The plot shows a clear peak at small values of mass $m\le0.005$, which is also an indication of the phase transition. The critical value of $\beta_c$ determined from the position of the peak is slightly larger and decreases when the mass decreases. This behaviour is expected, because it is well-known that nonzero mass shifts the position of the transition to larger values of $\beta$.

\begin{figure}[t]
\centering
\includegraphics[width=9.0cm,angle=0]{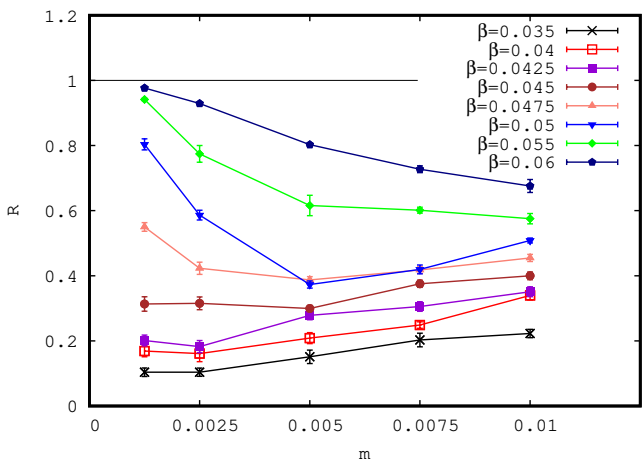}
\caption{Logarithmic derivative of $R$ of the chiral condensate as a function of mass $m$ for different values of coupling constant $\beta$.}
\label{fig:riso}
\end{figure}

 In the Fig.~\ref{fig:riso} we plot the $R$ as a function of $m$ for different values of $\beta$. Taking into
account the properties of the $R$ discussed above one can conclude that for $\beta\ge0.0475$ the system has no gap,
while for $\beta\le0.0425$ the results indicate the formation of the gap in the chiral limit. It allows to estimate the critical value of the coupling $\beta=0.0450\pm0.0025$, which is in the agreement with the estimation of $\beta_c$ from the data for the condensate.

\begin{figure}[t]
\centering
\includegraphics[width=9.0cm,angle=0]{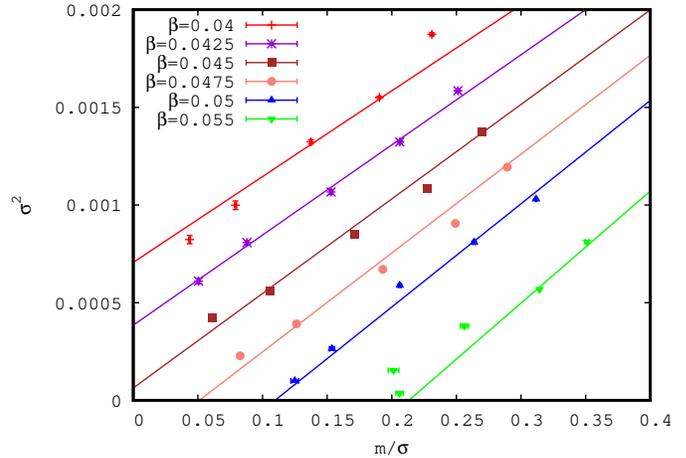}
\caption{The squared chiral condensate $\sigma$ as a function of $\frac{m}{\sigma}$.
Straight lines correspond to the fit of all points with EoS~(\ref{eq:eos}).}
\label{fig:fisheriso}
\end{figure}

To estimate the values of $\beta_c$ more precisely we fit the data with an equation of state (EoS) $m=f(\sigma,\beta)$. Motivated by studied of QED\cite{Gockeler:1996me} and graphene\cite{Drut:2008rg}, we applied the following equation of state:
\begin{equation}
    m X(\beta)=Y(\beta)f_1(\sigma)+f_3(\sigma),
\label{eq:eos}
\end{equation}
where one expands $X(\beta)=X_0+X_1(1-\beta/\beta_c)$, $Y(\beta)=Y_1(1-\beta/\beta_c)$ in the vicinity of critical $\beta_c$. For the left hand side we used classical critical exponents: $f_1(\sigma)=\sigma$, $f_3
(\sigma)=\sigma^3$. Such EoS can be easily visualized if one plots $\sigma^2$ as a function of $m_0/\sigma$ for various values of $\beta$ (Fisher plot). The resulting dependence $\sigma^2(m_0/\sigma)$ form straight lines, crossing the origin at $\beta_c$. This Fisher plot is presented in the Fig.~\ref{fig:fisheriso}. The deviations from straight lines might be attributed to finite volume effects or to non-classical critical exponents. The fit of the data in the vicinity of transition by Eq.~(\ref{eq:eos}) is given by the straight lines on the same Figure. Using this fit we obtained $\beta_c=0.04549(6)$. The presented error is only statistical. This value of $\beta_c$ corresponds to the critical coupling $\alpha_{eff}^c=1.749(2)$, which is close to the results obtained within the ladder approximation\cite{Gonzalez:2015tsa}, where critical coupling was found to be equal $\alpha_{eff}^c=1.8660$.

\begin{figure}[t]
\centering
\includegraphics[width=9.0cm,angle=0]{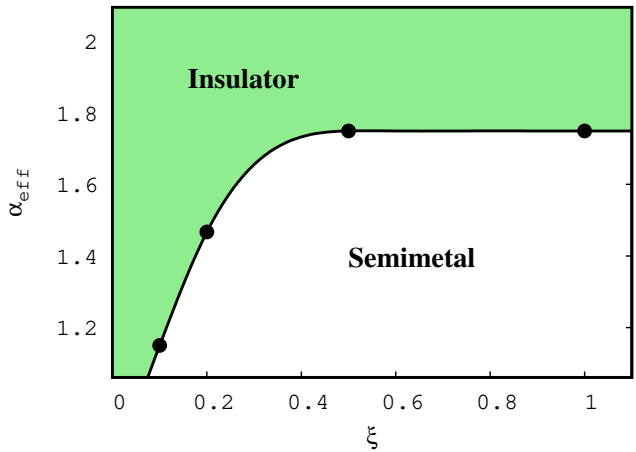}
\caption{The dependence of the critical coupling constant $\alpha_{eff}^c$ on the Fermi velocity anisotropy $\xi$. For $\alpha_{eff}>\alpha_{eff}^c(\xi)$ the system is in the insulator phase. Smaller values of $\alpha_{eff}<\alpha_{eff}^c(\xi)$ correspond to the semimetal phase. Statistical errors are smaller than data points. Lines are to guide the eyes.}
\label{fig:phase}
\end{figure}

Having accomplished the study of isotropic Dirac semimetals we proceed to the anisotropic case
which is parameterized by the parameter $\xi_3=\xi<1$ ($\xi_1=\xi_2=1$). The study
was conducted on the lattice $20^4$ using the procedure described above
for the value of the $\xi=0.1, 0.2, 0.5$. For these values of the $\xi$
the Figures \ref{fig:sigmaiso}-\ref{fig:fisheriso} are similar to that
for the isotropic case. For this reason we don't show them here.
We have found the following values of the critical $\beta_c$:
$\alpha_{eff}^c=1.762(3)$ for the $\xi=0.5$, $\alpha_{eff}^c=1.467(10)$ for the $\xi=0.2$ and $\alpha_{eff}^c=1.150(8)$
for the $\xi=0.1$.
So one sees that at $\xi=0.5$ the $\beta_c$ is practically the same as that at $\xi=1$.
For the values $\xi\leq0.2$ the value of the $\beta_c$ quickly increases with the decreasing
of the $\xi$. Tentative phase diagram is shown in Fig.~\ref{fig:phase}.

Note that the parameter $\xi$ effectively controls the dimension of the system under study. In the isotropic case $\xi=1$ the system is 3-dimensional. At the $\xi=0$ the system is similar to the stack of 2-dimensional sheets with Fermi velocity $v_{\parallel}$.
From quantum mechanics one may expect that the critical coupling for the 2D system is smaller than that for the 3D system, what is in agreement with the phase diagram in the Fig.~\ref{fig:phase}.

Detailed analysis of final volume effects requires considerable
computational resources and it will be done in a separate study. However,
in order to estimate the volume dependence of our results we carried out lattice simulation
of Dirac semimetals on the lattice $24^4$ for the asymmetries $\xi=0.1$ and $\xi=1$.
For the $\xi=0.1$ the critical coupling increases by 5\% and for the $\xi=1$ the critical
coupling increases by 10\%. So this shows that the volume dependence will not change our results
dramatically.

According to the experimental results the effective coupling constants for the Dirac semimetals
Na$_3$Bi, Cd$_3$As$_2$ are $\alpha_{eff}\simeq7$, $\alpha_{eff}\simeq2$ correspondingly.
Analysis carried out in this paper implies that Na$_3$Bi and Cd$_3$As$_2$ are deep in the insulator phase
what contradicts to the experiments. So, our paper rises very important question of the
theory of Dirac semimetals: why such
strong interaction in Dirac semimetals does not lead to dynamical generation of energy gap in the fermion
spectrum? A possible resolution of this puzzle is that in real world the
interaction potential is screened by bound electrons, what was not accounted in our study.
This mechanism is similar
to that observed in graphene \cite{Ulybyshev:2013swa}, where bound
electrons screen the interaction potential at small distances and shift the position
of the phase transition. Another possible explanation is that due to the renormalization effects
strong interaction can considerably modify the basic parameters of the theory.  
Although the study of different explanation of the raised question is very important
it is beyond the scope of this paper.

{\bf Conclusions.} In this paper the phase diagram of Dirac semimetals was studied within lattice Monte-Carlo simulation.
In particular, we concentrated on the dynamical chiral symmetry breaking which results in semimetal/insulator
transition. We measured the chiral condensate and the susceptibility of the chiral condensate for different values of the fermions mass,
effective coupling constant and the anisotropy of the Fermi velocity. Using these measurements we determined the values of the
critical coupling constant of the semimetal/insulator transition for different values of the anisotropy of the Fermi velocity.
This measurement allowed us to draw tentative phase diagram of Dirac semimetals.

It turns out that within the Dirac model with Coulomb interaction both
Na$_3$Bi and Cd$_3$As$_2$ known experimentally to be Dirac semimetals
would lie deeply in the insulating region of the phase diagram. It probably
shows a decisive role of screening of the interelectron interaction in real
materials, similar to the situation in graphene.

\section*{Acknowledgments}

The authors would like to express their gratitude to M.A.Zubkov, who drew their attention to the problem considered in this paper.
The authors thank I.A.~Shovkovy, Z.V.~Khaidukov for useful discussions and comments.
The work of MIK was supported by Act 211 Government of the Russian
Federation, Contract No. 02.A03.21.0006. The work of VVB and AYK, which consisted of
numerical simulation and the determination of the critical coupling constants at different Fermi velocity asymmetries,
was supported by grant from the Russian Science Foundation (project number 16-12-10059).
Numerical simulations were carried out on GPU cluster of NRC Kurchatov Institute.

\end{document}